\documentclass[%
preprintnumbers,
 amsmath,amssymb,
]{revtex4-1}

\usepackage{graphicx}
\usepackage{dcolumn}
\usepackage{bm}
\usepackage{mathtools}
\usepackage{amsmath}
\usepackage{amsmath,mathtools,amssymb}

\begin{document}

\preprint{APS/123-QED}

\title{Mapping the Design Space of Photonic Topological States via Deep Learning}

\author{Robin Singh}
\email{robinme@mit.edu}
\affiliation{Department of Mechanical Engineering, MIT.nano, \\
Massachusetts Institute of Technology}
\author{Anuradha Murthy Agarwal}%
\affiliation{Department of Materials Science and Engineering,\\ Microphotonics Center, Materials Research Laboratory,\\Massachusetts Institute of Technology
}
\author{Brian W Anthony}
\email{banthony@mit.edu}
\affiliation{Department of Mechanical Engineering, MIT.nano, \\
Massachusetts Institute of Technology}
 \altaffiliation[Also at ]{Department of Mechanical Engineering, Massachusetts Institute of Technology}



\date{\today}

\begin{abstract}
Topological states in photonics offer novel prospects for guiding and manipulating photons and facilitate the development of modern optical components for a variety of applications. Over the past few years, photonic topology physics has evolved and unveiled various unconventional optical properties in these topological materials, such as silicon photonic crystals. However, the design of such topological states still poses a significant challenge. Conventional optimization schemes often fail to capture their complex high dimensional design space.  In this manuscript, we develop a deep learning framework to map the design space of topological states in the photonic crystals. This framework overcomes the limitations of existing deep learning implementations. Specifically, it reconciles the dimension mismatch between the input (topological properties) and output (design parameters) vector spaces and the non-uniqueness that arises from one-to-many function mappings. \\

We use a fully connected deep neural network (DNN) architecture for the forward model and a \textit{cyclic convolutional neural network (cCNN)} for the inverse model. The inverse architecture contains the pre-trained forward model in tandem, thereby reducing the prediction error significantly. 

\end{abstract}

\maketitle


\section{Introduction}
With the advent of topological phases of matter in electronic materials, the exploration of band properties in photonic materials have received a huge impetus. A photonic system's topological effects emerge from the classical behavior of light waves, which results in different topological phases that manifest as geometrical properties of the system. Conventionally, photonic crystals (PhCs) are a useful class of structures for understanding topological states and their "high order" physics. They are engineered structures that have periodic arrangements of two or more materials varying significantly in their dielectric constants [2,3], and are analogous to electronic crystals that are used to control the electronic properties in condensed matter physics [1]. Depending on the extent of periodicity in space, they can be 1-D, 2-D, or 3-D photonic crystals. They have been front and center in photonics research for more than a few decades now, due to their immense potential to guide and propagate light. Well-designed nanostructures in the photonic crystals can support unique properties that allow effective photon transport in photonic integrated circuits [4]. Topological photonic crystals guide light in unconventional ways, enabling entirely new ways to route information for communication and computing purposes. Hence, they are used for various photon manipulative applications, to reflect and trap light, thereby forming reflective surfaces, waveguides, and resonant cavities. \\

With advances in silicon photonics being widely used in telecommunications and optical sensors, PhCs are being explored for developing optical components such as optical filters, waveguide bends, and power splitters, all with improved performance. Dielectric PhCs overcome some conventional engineering challenges by offering weak frequency dependency [3,5,6]. For instance, metallic components were traditionally used to guide, reflect and trap light. But, they suffer from high dispersive losses due to their complicated electronic properties that depend on the frequency, unlike dielectric PhC counterparts. \\

Mainly, when the wavelength of the light is comparable to the periodicity of the crystal, various exciting phenomena have been reported [7]. Of many vital properties of topological PhC, the most interesting is the formation of a photonic bandgap (PBG) structure. PBG refers to the phenomena where photons within a range of frequencies are not allowed to propagate through the PhCs. We can design and construct PhCs with a specified band gap allowing light to propagate in a specified direction with specified frequencies. It is a fundamental property of PhCs that gave rise to industrial applications of lossless waveguides, light-emitting diodes (LEDs), and high Q factor resonant cavities. Fig. 1 represents the band structure of a typical 1D photonic crystal [8-11]. \\

 Researchers have relied on long-range ordered structures based on the material crystal structure, or quasi-structure, to engineer PhC devices with desired properties. However, PhCs are sensitive to material defects, thus making them difficult to fabricate with this technique. Recent advances in nanoscale optical phenomena, computational tools and fundamental understanding of PhCs, have enabled the design of novel complex structures. Photonic topology physics has evolved leading to unconventional optical properties in these topological materials, whose states are often high-ordered and require tighter control of the geometry of structured material, posing a significant design challenge [12,13]. \\

Currently, photonic devices are designed using conventional optimization procedures that start with random design parameters, compute its electromagnetic response using methods of Finite Element (FE) analysis, Finite Difference Time Domain analysis (FDTD). The calculated response is compared with the targeted response, evoking a structural update in the design, and the process iterates 1000s of times until a stopping threshold is reached. Notable examples that are built on it are level-set methods, and adjoint methods [6]. These computational techniques have numerical drawbacks. First, they are computationally expensive, as a slight modification in the design triggers a long iterative procedure every time. Second, they get prohibitively slow as the size and complexity of the problem grow. Moreover, as the design space grows high in its dimension, the optimization fails.\\ 

Machine Learning methodologies are becoming prevalent methods to perform a thorough search of the design space efficiently. More recently, supervised learning methods such as deep neural networks are getting a lot of attention. In a neural network architecture, it is possible to represent complex mapping functions according to the \textit{universal approximation theorem} [12]. The different layers of neurons in the network enable learning of complex mathematical functions and hence enable the development of forward and inverse models in estimating the properties of topological PhC devices.  The forward and inverse models are defined as shown in Fig. 3a.  Several studies have shown the DNN that connects more generic PhC structures to their EM response. However, current challenges of the DNN implementations are (1) Dimensional mismatch between input and output space (2) Inverse problem estimation due to implicit conflicting instances of one-to-many mapping functions. \\

In this manuscript, we explore machine learning-based methods to design photonic crystal structures with targeted topological properties.  We consider a simple 1D dielectric photonic crystal structure that supports non-trivial topological properties. The crystal has periodicity in the material properties (in terms of permittivity) along the x-direction. We present 'forward' and 'inverse' model estimation using trained DNN and CNN architecture. Through our demonstration, we thereby attempt to solve the presented challenges i.e., dimensional mismatch in mapping the topological PhC design space and inverse problem estimation.  It should be noted that training a forward model can be done using a standard process as the mapping function is one-to-one in nature.  However, the inverse model estimation can be tricky. The problem arises due to the ill-posed nature of the inverse scattering problem, which states that the same electromagnetic response can be created by two different designs of the PhCs. We overcome this challenge by using a \textit{cyclic convolutional neural network} architecture that helps to reduce the estimation errors significantly. The presented formalism can be easily scaled to more complex structures in 2D and 3D geometries. 

\subsection{Related Work}
Yashar Kiarashinejad et al. demonstrated a new approach based on deep learning (DL) techniques for analysis, design, and optimization of electromagnetic (EM) nanostructures [14]. Their model ensured that physically viable solutions are obtained from the trained inverse model of the topological structure. However, the physical constraints on the inverse solutions get tricky when we deal with a more complex topological structure. In another report, Zhang et al. presented neural networks that could successfully learn topological invariants for topological band insulators [15]. Adler et al. presented a neural network-based method that could learn the ill-posed inverse model through prior knowledge [16]. Liu et al. demonstrated the inverse design of nano-photonic structures using DNN and overcame the ill-posed nature of inverse problem by using a tandem network of inverse and forward models [12,17]. Goshal et al. developed neural network-based methods to design EM wave generator antenna designs [18]. In a recent study, Peurifoy et al. presented the possibility of estimating the EM scattering profile from photonic nanoparticles with high accuracy [19]. Liu et al. developed a free open source EM solver for periodic structure [20]. 

\section{Results and Discussions}

\subsection{1D Photonic Crystal and Topological Properties}
We leverage the role of symmetry in estimating the models for our 1D crystal (Fig. 1). It plays an essential role in training our neural network in two ways. First, the symmetry of the crystal needs to be taken into account while developing the training data. Ideally, to allow the neural network to learn a general rule, the training data needs to be free from constraints arising from symmetry bias. Second, based on the symmetry in our crystal structure, we choose an architecture of the neural network that is compatible with the symmetry of targeted physics behind the model. To take this into account, we calculate the band structure for one half of the PhC and mirror flip it to obtain the entire crystal (Fig. 1b).  

\subsection{Formulation of the Machine Learning Problem}

Having understood the topological properties and design parameters of the PhCs (shown in Fig. 1), we provide the mathematical formulation of 'forward' and 'inverse' model to define the machine learning problem. For the given design vector space of the photonic structure, $D_i$, we obtain the forward model, through a mapping function defined as, \begin{equation}
B = \mathbb{F}({D_1} = {L_1},{D_2} = {L_2},{D_3} = {L_3},{D_4} = {L_4})
\end{equation}
here, $B$ is the observed output space, in our case, it is the band gap structure. $\mathbb{F}$ is the one-to-one mapping function. Forward model estimation deals with predicting the most faithful function, $\hat{\mathbb{F}}$. A common approach is to define an objective function as,  
\begin{equation}
\underset{ D_i \in D}{\text{minimize}} [\ell(\hat{\mathbb{F}}(D_i),g)]
\end{equation}
where, $\ell$ is the suitable affine transformation of the data, defined as 
\begin{equation}
    \ell=\mid \mid \hat{\mathbb{F}}(D_i)-g\mid \mid
\end{equation}

where in, $\hat{\mathbb{F}}$ is the predicted forward function and $g$ is the ground truth ($B$, band gap structure). Hence, machine learning problem for the forward model is to find the best estimator of $\hat{\mathbb{F}}_\Theta$   given the observations (training data). $\hat{\mathbb{F}}_\Phi$ is the parameterized using $\Phi \in Z$  where the optimal $\Phi$ is obtained from the training data.\\

On the other hand, an inverse problem is reconstructing a design space, $D_i$ from the data, $B_i$ shown as,
\begin{equation}
{D_i} = {\mathbb{F}^{ - 1}}({B_1},{B_2},{B_3},{B_4},..{B_n})
\end{equation}

Where, $D$ (design parameters) and $B$ (band gap structure) are the topological vector spaces, $\mathbb{F}^{-1}$ is the mapping function between them. Often, this function cannot be analytically inverted. Hence, we instead use pseudo inverse, $\mathbb{F}^{-1}$. Finding pseudo inverse is an ill-posed problem, i.e., a solution (if its exists) is unstable with respect to inputs. Hence, the classical regularizer defined for the forward model does not work well.  Machine learning is a promising avenue for non-linear approximation that can be applied for such pseudo-inverse function, ${\hat{\mathbb{F}}_\Theta }^{ - 1}(g = {B_i}) \approx {D_i}$ . Similarly to the forward model, we parametrize the pseudo inverse operator, $\hat{\mathbb{F}}_\Theta$  by $\Theta \in Z$, where $Z$ is a suitable parameter space. The parameter approximation is obtained through neural network that learns the ‘optimal’ $\Theta$  from the training data. 
\subsection{Deep Neural Network for Forward Model}

As defined in equations-1, 2 and 3, the forward model approximates the mapping function, $\mathbb{F}$  between D and B vector space through $\hat{\mathbb{F}}$ . We use neural network based regression to develop the multi-dimensional model. It consists of computational layers that are connected through an activation function.  Fig. 2a shows the NN architecture of the forward model used in the study. More details on the activation function, loss function can be found in the Methods section. With rigorous architectural experimentation, we find that 4-8-16-32-55 fully connected layer-based NN outperforms others for the training dataset of 50000 unique datasets.  Fig. 3c shows the convergence of the optimization function represented through decreasing mean square error (MSE). After 500 iterations, we find that the test error reduced to as low as 0.0081 with almost zero-approaching slope of the convergence plot. Fig. 3d compares the predicted band structure for a test case with the ground truth. Hence, we conclude that NN is able to estimate the band structure faithfully.  Further, this model can be scaled to take in account other design parameters, such as di-electric constants (${\varepsilon _1}$, ${\varepsilon _2}$) of the layers, while predicting the band structures. 
\begin{widetext}
\begin{figure}
  \includegraphics[width=\linewidth]{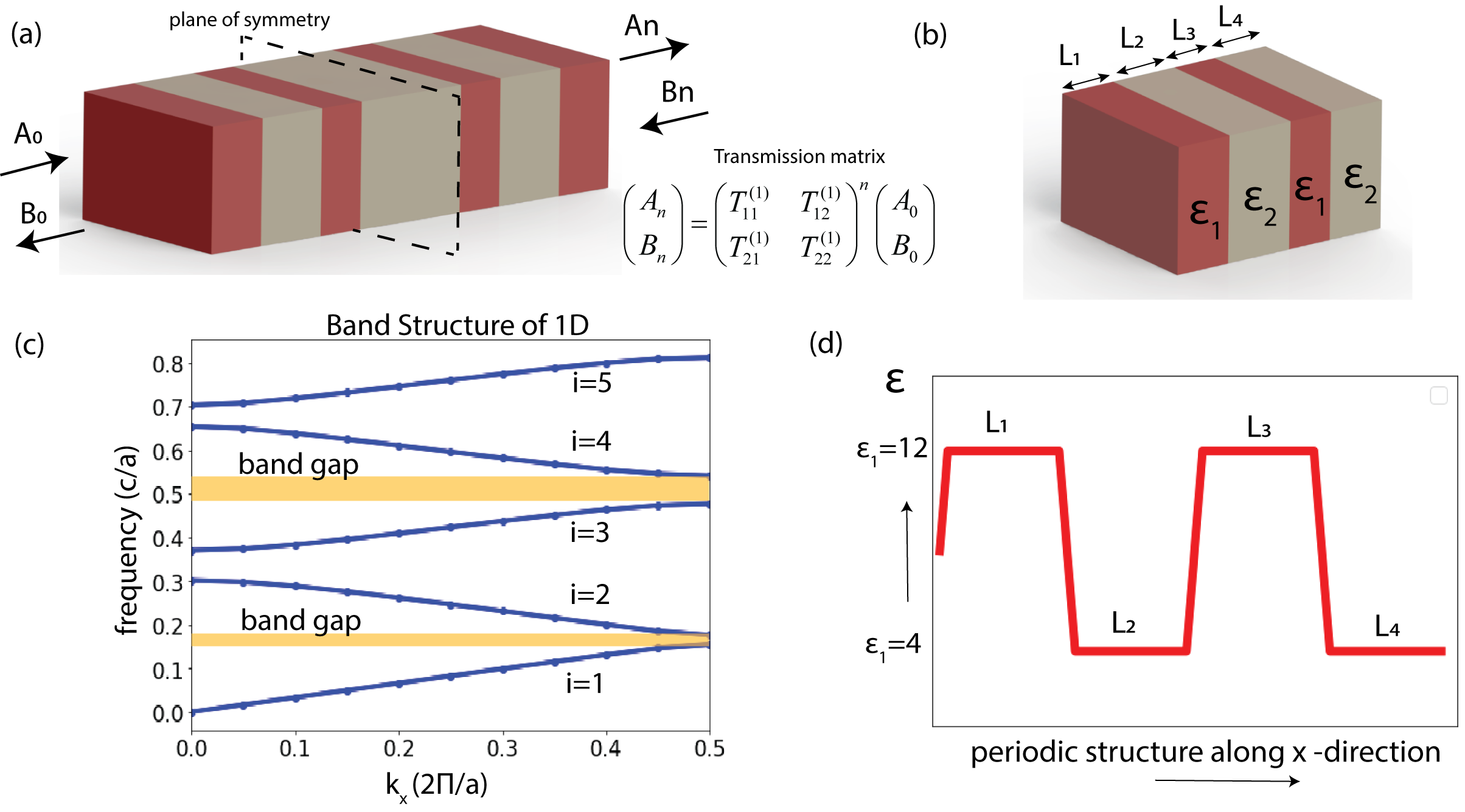}
  \caption{Typical 1D photonic crystal consisting of periodic structures with different dielectric constants. Input and output wave forms are related to each other through the transmission matrix, T. (b) Exploiting the symmetry of the 1D crystal, we consider one half of the structure. (c) Band structure of the PhC when $L1=L2=L3=L4=0.25a$, where a is the half length of the crystal. (d) The dielectric distribution in the crystal structure. }
  \label{fig:boat1}
\end{figure}
\end{widetext}
\subsection{Cyclic Convolutional Neural Network for Inverse Model}
As shown in Fig. 3a, an inverse problem refers to reconstructing the design parameters that characterize the system under investigation from the bandgap structures.  The pseudo-inverse operator 
${\mathbb{F}_\Theta }^{ - 1}$ is approximated through neural network architecture. During our initial trials to invert the model, we find that a convolutional neural network (CNN) outperforms the fully connected DNN.  They tend to have theoretically better performance compared to the fully connected DNN with the higher relative efficiency of their local architecture. With their filter definitions, they can pick up complex features in the input band structure. We use 1 convolutional layer connected to 3 fully (320-64-32-4) connected layers. Fig. 2a shows the CNN architecture. More details on the other features of CNN (activation function, filter size, max pooling layer) can be found in the Methods section of the manuscript.  Since more than one set of design parameters can result in similar band structure; we suspect that there would be conflicting instances while training the NN. It might result in poor convergence in the model.\\

To overcome these issues, we consider the algorithm, as explained in Fig. 2c, where we validate the design parameter generated from the CNN with the forward model. In this procedure, the tentative solution ($D_i$) obtained from CNN is fed into the forward model that uses the predicted design parameters to generate the band structure ($B_j$). The algorithm then compares $B_j$ with $B_i$ (input of the CNN) and performs an update on the model. It continues until we reach the desired accuracy level of the prediction. The neural network equivalence of the algorithm is shown in Fig. 2d.  Such network is often referred to as ‘Cyclic Convolutional Neural Network (cCNN)’ as the predicted output is compared against the input to update the weights in the layers. The tandem network has CNN input connected to the pre-trained forward model, F.  During the training, we freeze weights of the pre-trained model and only change the weights of the CNN during. After 500 iterations, we find that testing error decreases significantly to a value of 0.014/sample. The tandem nature of the model ensures that the model obtained is indeed viable and helps in avoiding them multivalued degeneracy to an extent.   Fig. 3f and 3g compare the predicted design parameters with the ground truth. Since we see that there is some considerable difference in the estimated design parameters, we cross-validate the NN performance by plugging in the design parameters to MPB simulations (details in the Methods section) and obtain the band gap structures. We find that the bandgap structure matches perfectly with the input to ‘cCNN’ confirming many-to-one mapping of the function. 

\section{Outlook}
We present a deep learning framework to design the topological states of PhCs. With the advent of more complex photonic structures and high order topological physics, conventional methods are not sufficient to capture the high order design space and often fail to produce faithful results. The present framework, on the other hand, represents a holistic approach towards the design. We present forward and ‘inverse’ model estimation to design simpler yet non-trivial topological states in 1-D PhCs. Currently existing machine learning based methods often fall short in developing the inverse model for the design due to its ill-posed nature. We overcome this issue in training our inverse model using cyclic network architecture to produce physically relevant parameters and avoid the network inconsistency due to one-to many mapping. \\
The proposed framework can be extensively applied well beyond the example presented in the manuscript. The method can be easily scaled to develop a more complicated photonic crystal structure with 2D and 3D geometries. It can be adapted to perform topological science studies in quantum technologies, polaritonics and ultra-cold atomic physics. Further work in developing such photonic structures can result in novel sensing modalities for a variety of applications [21,22]. 

\section{Methods}

\subsection{MPB simulations}
The electromagnetic response of the PhC device is obtained using MPB simulations. It is an open-source software package for computing the band structures, or dispersion relations, and electromagnetic modes of periodic dielectric structures. We use serial computers to generate the data on band gaps for different design parameters of the 1D material stack. We use two materials with a significant difference in their dielectric properties, one with permittivity of 12 and the other with 4 to generate the crystal device. The ambient environment for the simulation is considered to be air with permittivity of 1.  We use the simulation domain of unit length (a) in the x-direction. We then define a range of $k$ (wave-number) values in the irreducible Brillouin zone to obtain the band structures of the photonic device. We defined about 9 k-data points between $0<k<0.5\pi/a$  and use a  resolution of 32 to obtain the first 5 band structures. 
\subsection{Data Generation, Formatting, and Preprocessing}
We use the random parameter generator to define thickness of 4 different layers. The code is constrained to generate 4 numbers that add to 1. The combination of these 4 layers is along the x direction. MPB library is installed on Python to automate the data generation.  We define the MBP environment and run it to generate 50000 training data set and 1000 test data set. The datasets are stored as .text files that are later loaded in Pytorch based neural network code.
\subsection{Training Procedure}
Neural Network is implemented through the Pytorch library on Python.  As such, we begin with defining the neural network architecture using the hidden layers and units. We test with different architectures to optimize the estimation error. The output from the hidden units is passed through the activation function. It is followed by defining the loss function and optimizer to estimate the values of the parameters. Once the appropriate functions are defined in the software, we perform gradient descent based optimization to estimate the model parameters. We implement the stochastic gradient descent (SGD) based optimizer to ensure the faster convergence in our training iterations. 
\subsection{Hyperparameters, Loss, and Activation Function}
The most important hyper parameters that are optimized in the experiments are batch size for SGD, number and size of hidden layers in NN, loss function and activation function. We find that smaller batch size results in smaller training and test error. Hence, we use the batch size of 50. Forward model estimator, F is composed of 3 hidden layers and the inverse model, G has one convolutional layer connected to 3 fully connected layers.  Following the standards, we use ReLU as the activation function for hidden layers. However, for the inverse model, G, the output layer uses sigmoid as the activation layer. This ensures the output layer values are bounded by 1 (restriction imposed by our training data set). We use ‘mean square loss (MSE)’ as the loss functions for the forward ( $\frac{1}{2}\sum\limits_{i = 1}^n {{{\left( {{B_{predicted}} - {B_{true}}} \right)}^2}} $ ) and inverse model ($\frac{1}{2}{\sum\limits_{k = 1}^n {\left( {{B_j} - {B_i}} \right)} ^2}$ ) as well. 

\begin{acknowledgments}
Authors would like to thank Manish Singh, a member of Laboratory of Financial Engineering \& CSAIL MIT for the useful insights in machine learning methods. Authors would also thank Alex Benjamin, a member of Device Realization Laboratory for the useful insights in improving the manuscript.   
\end{acknowledgments}

\section{Authors Contributions}
R. S. performed the integrated photonic modeling, and machine learning experiments. A.A and B.W.A provided guidance, technical discussions and advised the research. All authors reviewed and edited the manuscript. 

\section{Data Availability}
MPB script files and any other accompanied codes used for modeling and training of the meta-surface are available from the corresponding authors upon reasonable request. 

\section{Conflict of Interest}
The authors declare that they have no conflict of interest.


\section{References}
1.	B.-Y. Xie, H.-F. Wang, X.-Y. Zhu, M.-H. Lu, Z. D. Wang, and Y.-F. Chen, Opt. Express 26, 24531 (2018).\\
2.	L. Lu, J. D. Joannopoulos, and M. Soljačić, Nat. Photonics 8, 821 (2014).\\
3.	J. D. Joannopoulos, S. G. Johnson, J. N. Winn, and R. D. Meade, 2nd ed. (Princeton University Press, 2008).\\
4.	T. Ozawa, H. M. Price, A. Amo, N. Goldman, M. Hafezi, L. Lu, M. C. Rechtsman, D. Schuster, J. Simon, O. Zilberberg, and I. Carusotto, Rev. Mod. Phys. 91, 15006 (2019).\\
5.	Y. Yang, Y. Yamagami, X. Yu, P. Pitchappa, J. Webber, B. Zhang, M. Fujita, T. Nagatsuma, and R. Singh, Nat. Photonics (2020).\\
6.	R. Singh, Y. Nie, A. M. Agarwal, and B. W. Anthony, ArXiv preprint (2019).\\
7.	S. Peng, N. J. Schilder, X. Ni, J. Van De Groep, M. L. Brongersma, A. Alù, A. B. Khanikaev, H. A. Atwater, and A. Polman, Phys. Rev. Lett. 122, 117401 (2019).\\
8.	A. J. Garcia-Adeva, New J. Phys. 8, (2006).\\
9.	N. Parappurath, F. Alpeggiani, L. Kuipers, and E. Verhagen, Sci. Adv. 6, 1 (2020).\\
10.	R. Singh, D. Ma, L. Kimerling, A. M. Agarwal, and B. W. Anthony, ACS Sensors 4, 571 (2019).\\
11.	R. Singh, P. Su, L. Kimerling, A. Agarwal, and B. W. Anthony, Appl. Phys. Lett. 113, 231107 (2018).\\
12.	D. Liu, Y. Tan, E. Khoram, and Z. Yu, ACS Photonics 5, 1365 (2018).\\
13.	S. Li, H. Lin, F. Meng, D. Moss, X. Huang, and B. Jia, Sci. Rep. 8, 1 (2018).\\
14.	Y. Kiarashinejad, S. Abdollahramezani, and A. Adibi, npj Comput. Mater. 6, 1 (2020).\\
15.	P. Zhang, H. Shen, and H. Zhai, Phys. Rev. Lett. 120, 66401 (2018).\\
16.	J. Adler and O. Öktem, Inverse Probl. 33, (2017).\\
17.	Z. Liu, D. Zhu, S. P. Rodrigues, K. T. Lee, and W. Cai, Nano Lett. 18, 6570 (2018).\\
18.	G. Gosal, E. Almajali, D. McNamara, and M. Yagoub, IEEE Antennas Wirel. Propag. Lett. 15, 1483 (2016).\\
19.	J. Peurifoy, Y. Shen, L. Jing, Y. Yang, F. Cano-Renteria, B. G. DeLacy, J. D. Joannopoulos, M. Tegmark, and M. Soljačić, Sci. Adv. 4, 1 (2018).\\
20.	V. Liu and S. Fan, Comput. Phys. Commun. 183, 2233 (2012).\\
21. Robin Singh, M.S. Thesis, Massachusetts Institute of Technology (2018).\\
22. R. Singh, A. Agarwal, and B. W. Anthony, 1124010, 35 (2020).

\begin{figure}
  \includegraphics[width=\linewidth]{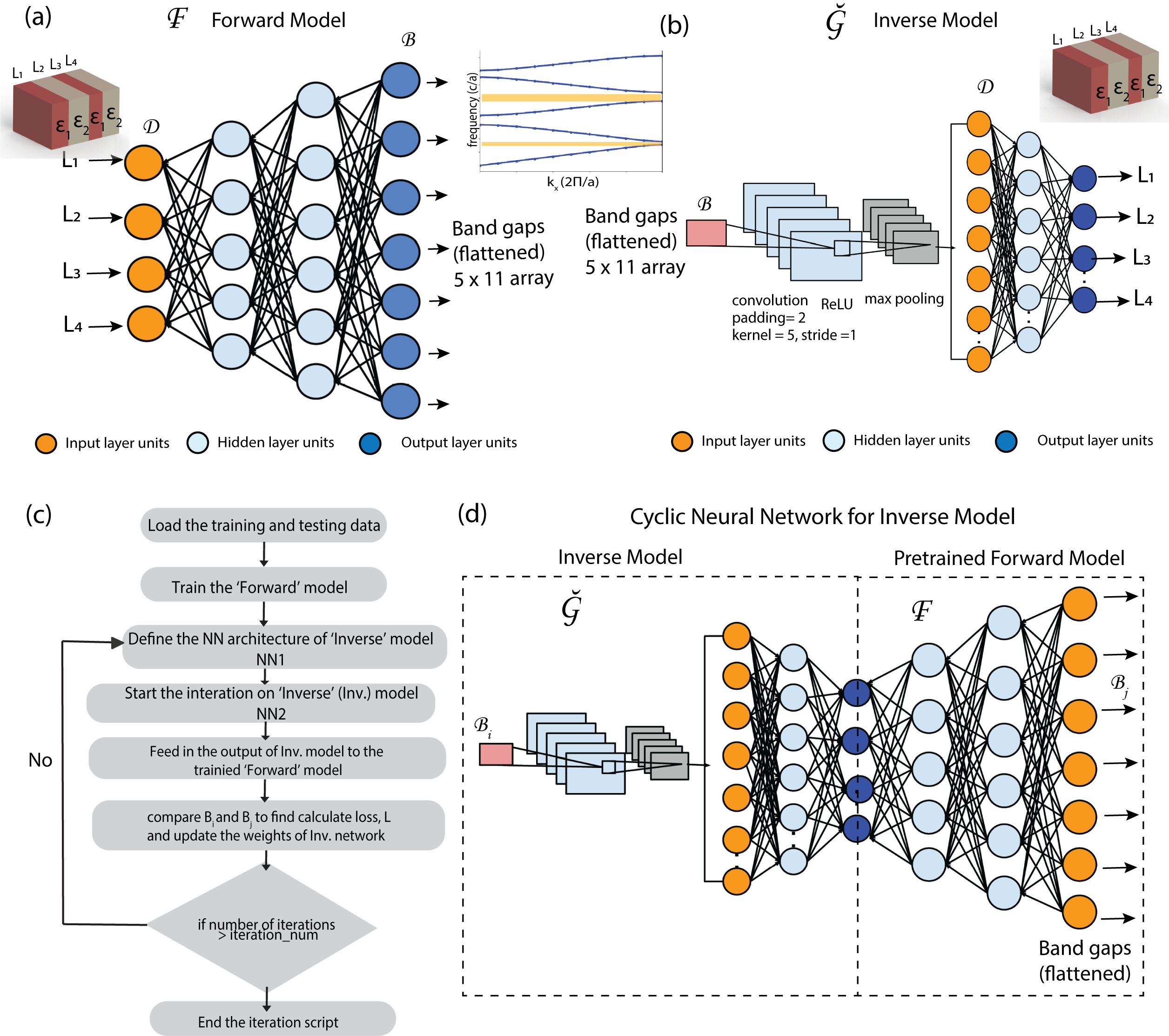}
  \caption{Deep neural networks for forward and inverse models. The forward estimator is given by $\mathbb{F}$  and the inverse estimator is given by $G$. For the forward model, we use 4-8-16-32-55 fully connected layers with design parameters of L1, L2, L3 and L4 as inputs and the output is 55 element flattened band gap structure. The output is post processed approximately to plot the band structure. (b) Inverse model is composed of a convolutional layer connected to 3 fully connected layers. (c) Algorithm process that feeds into the output the inverse model, G to the pre-trained forward model. The tandem network outperforms the simpler inverse model, $G$.   }
  \label{fig:fig2}
\end{figure}

\begin{figure}
  \includegraphics[width=\linewidth]{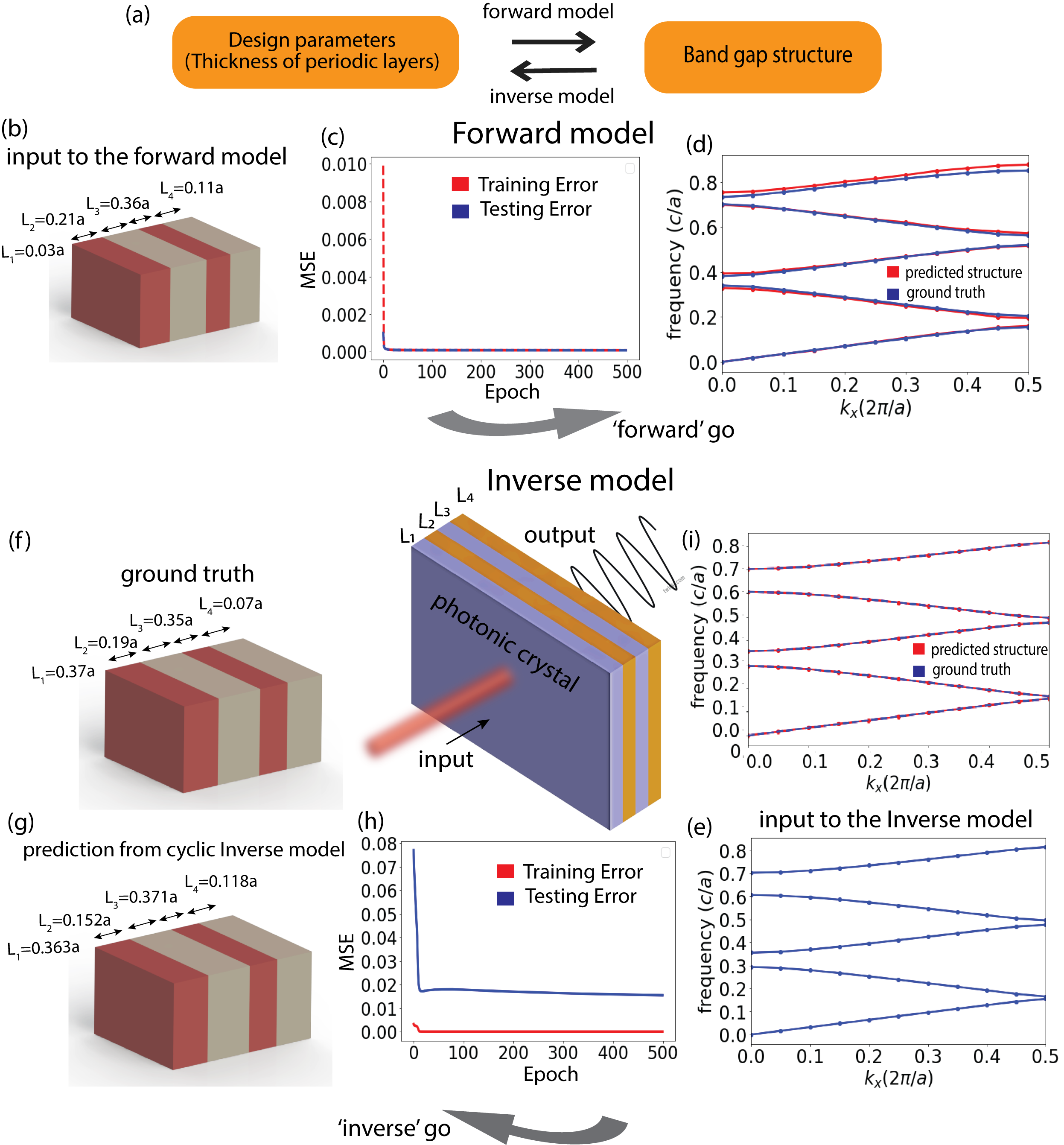}
  
  \caption{Performance of Deep neural networks for forward and inverse models. (a) Definition of the ‘Forward’ and ‘Inverse’ model to characterize the topological states in 1D photonic crystal. (b) Input to the forward model; Design parameters of the photonic crystal. (c) Convergence of ‘Training’ and ‘Testing’ error with the number of iterations. (d) Predicted band structure of the crystal and its comparison with the ground truth. (e) Input to the Inverse model; Band gap structure. (f) Ground truth of the design parameters (g) Predicted values of the design parameters from ‘Cyclic Inverse’ model. (h) Convergence of MSE for ‘Training’ and ‘Testing’ errors with number of iterations. (i) Cross-validation of the predicted output from ‘Inverse’ model by comparing the generated band gap structure with the input band structure.}
  \label{fig:fig3}
\end{figure}
\end{document}